# CHIRALLY INVARIANT AVATAR IN A MODEL OF NEUTRINOS WITH LIGHT CONE REFLECTION SYMMETRY

Alan Chodos*

*Haseltine Systems, 2181 Jamieson Ave., Suite 1606*
*Alexandria, VA 22314*

**Abstract**: In previous work we developed a model of neutrinos based on a new symmetry, Light Cone Reflection (LCR), that interchanges spacelike and timelike intervals. In this paper we start with the four-dimensional model, and construct a two-dimensional avatar that obeys the same equations of motion, and preserves both the light-cone reflection symmetry and the chiral symmetry of the original theory. The avatar also contains the interaction that rendered the four-dimensional model gauge invariant. In an addendum, we make some remarks about how to determine the scalar field that enters into the definition of the LCR-covariant derivative.

*chodos@aps.org

**Introduction**

This is the fourth in a series of papers [1] investigating Light Cone Reflection (LCR) as a possible symmetry of neutrinos. The general idea, and motivation, is that this symmetry, which interchanges spacelike and timelike intervals, will lead to a spectrum of neutrinos that includes both subluminal and superluminal varieties, building on a possibility suggested more than 30 years ago [2].

The desirability of such a possibility is open to question. Presumably superluminal propagation leads to violations of causality, a cherished principle whose sanctity is comparable to that of conservation of probability or a Hamiltonian that is bounded below. Indeed, there are those [3] who would argue that pure thought alone is enough to rule such causality violation out. However, the history of physics is littered with the corpses of proofs based on lines of reasoning with which Nature happened not to agree. The speed of neutrino propagation is, and must finally be, a question for experiment to decide.

One may think that the collapse of the claim made by the OPERA collaboration some years ago [4] confirmed experimentally that neutrinos do not travel faster than light. Up to a point that is the case. After correcting its initial, superluminal result, OPERA measured neutrino velocities consistent with the speed of light within certain error bars, whose width, however, did not rule out mass parameters, either sub- or superluminal, in the meV range that one expects for neutrinos.

In the first paper of this series, we introduced the notion of LCR, making use of the framework of Very Special Relativity (VSR) that was put forward about a decade ago by Cohen and Glashow [5]. Our theory, therefore, does not fully realize four-dimensional Lorentz symmetry, but comes close, being invariant under the maximal subgroup sim(2). In operational terms, we introduce a null vector $n^\mu$, and demand that the theory be invariant under all Lorentz transformations that either leave $n^\mu$ invariant or else scale it by a constant factor.

In paper II, we found a minimal realization of LCR, in which a scalar field, subject to a VSR-invariant constraint, was used to construct an appropriate covariant derivative. Including both Dirac and Majorana mass terms, we then found a transformation to "special coordinates", in which we determined the spectrum of the theory, showing that it was indeed symmetric between tachyonic and non-tachyonic species.

In paper III, we introduced gauge interactions. From previous work, we knew that not all the degrees of freedom appearing in the model were dynamical, and we were able to choose the particular form of the interactions so that the gauge fields coupled only to the dynamical degrees of freedom. In addition, in the special coordinate system the fields depended only on two of the four coordinates. Restricting ourselves to the non-interacting case, in the third paper we constructed an "avatar" of the four-dimensional theory, i.e. a set of fields in two dimensions, obeying first-order equations which when squared yielded the same second-order equations that determined the spectrum of the original theory. We quantized the avatar, and obtained the expected spectrum of tachyonic and non-tachyonic states.

In this paper, we extend the previous treatment in a number of ways. We find an additional kinetic term for the gauge fields, with all the right symmetries, that should be included in the theory; indeed, the original kinetic term vanishes for solutions of the equations of motion, and the behavior of the gauge fields is actually governed by the newly added term. In paper III, the avatar was simply postulated, and it was then verified that the correct spectrum was obtained. However, the avatar in that paper did not respect the chiral symmetry of the original four-dimensional model. Here we remedy this defect by constructing the avatar directly from the four-dimensional model. We find that the avatar previously studied contained only half the necessary degrees of freedom. In addition, this construction allows us to properly introduce the gauge interactions at the level of the avatar. We note the similarity of this interacting system to certain exactly solvable models in two dimensions.

**The Four-dimensional Theory**

As in previous work, we wish to construct a theory of fermions invariant under VSR, LCR and chiral symmetry. (As Cohen and Glashow originally pointed out, the intervention of the matrix $n \cdot \gamma$ allows for chiral symmetry even in the presence of mass terms). The LCR transformation is

$$x^\mu \to y^\mu = x^\mu - n^\mu x^2 / n \cdot x \quad , \text{ with } n_\mu n^\mu = 0,$$

which clearly satisfies $y^2 = -x^2$. We introduce a covariant derivative

$$D_\mu = \partial_\mu - \partial_\mu \phi \, n \cdot \partial,$$

where $\phi$ is a scalar field satisfying the constraint $n \cdot \partial \phi = 1$. Note that

$$n^\mu D_\mu = 0 \quad \text{and} \quad D_\mu \phi = 0.$$

Under LCR,

$$\phi(x) \to -\phi(y),$$

which both preserves the constraint (because $n \cdot \partial$ is odd under LCR) and insures that $D_\mu(x) \to D_\mu(y)$.

To implement LCR invariance, we introduce two fermion fields, $\psi^{(a)}$ and $\psi^{(b)}$, respectively even and odd under LCR. In addition, we follow Alvarez and Vidal [6] and add a pair of auxiliary fermion fields, $\chi$ and $\rho$, with $\chi$ even and $\rho$ odd under LCR. As discussed in paper III, we also include gauge fields $A_\mu$ and $B_\mu$. The Lagrangian is

$$L = L_\psi + L_{\chi\rho} + L_M + L_{int} + L_{AB},$$

where

$$L_\psi = i\overline{\psi^{(a)}}\gamma^\mu D_\mu \psi^{(a)} + i\overline{\psi^{(b)}}\gamma^\mu D_\mu \psi^{(b)},$$

$$L_{\chi\rho} = i[\overline{\chi}(n\cdot\partial)\rho + \overline{\rho}(n\cdot\partial)\chi],$$

$$L_M = i\{M_1\overline{\chi}\gamma^\mu n_\mu \psi^{(a)} + M_2\overline{\rho}\psi^{(b)} - h.c.\},$$

$$L_{int} = g\sum_{j=a,b}\left[i\overline{\psi^{(j)}}\gamma^\mu n\cdot\gamma\psi^{(j)}A_\mu + \overline{\psi^{(j)}}\gamma^\mu\gamma_5 n\cdot\gamma\psi^{(j)}B_\mu\right], \text{ and}$$

$$L_{AB} = K\left[F^A_{\mu\nu}F^{A\mu\nu} + F^B_{\mu\nu}F^{B\mu\nu}\right]\phi^{-2} + k\{[n\cdot\partial A^\mu][n\cdot\partial A_\mu] + [n\cdot\partial B^\mu][n\cdot\partial B_\mu]\}.$$

Here $F^A_{\mu\nu} = D_\mu A_\nu - D_\nu A_\mu$, and similarly for $B_\mu$. Under chiral transformations, $A_\mu$ and $B_\mu$ transform as a doublet. To make $L_{int}$ Hermitian, they are required to satisfy $n^\mu A_\mu = n^\mu B_\mu = 0$. The constants K and k can be chosen for convenience.

In paper II we included Majorana mass terms in $L_M$, but for the purposes of constructing gauge-invariant interactions we restrict ourselves to Dirac terms only, as we did in paper III. The mass parameters $M_1$ and $M_2$ may be complex. In paper III we allowed for different couplings $g_a$ and $g_b$ to $\psi^{(a)}$ and $\psi^{(b)}$, but here for simplicity we employ only a single coupling constant g.

The second term in $L_{AB}$, proportional to k, was missing in paper III, but it possesses all the necessary symmetries and should be included. As discussed in paper III, we have made a gauge choice that leaves a residual gauge symmetry for which the gauge parameters are annihilated by $n\cdot\partial$. Thus the term proportional to k is not only VSR, LCR, and chiral-invariant, but is invariant under the residual gauge symmetry as well.

When $n^\mu$ is scaled by a factor $\alpha$, the gauge fields A and B scale by $\alpha^{-1}$, which renders $L_{int}$ invariant, but requires a compensating factor of $\phi^{-2}$ in the first term of $L_{AB}$. In paper III we chose not to have A and B scale, so $\phi$ appeared in $L_{int}$ instead of $L_{AB}$.

## Equations of Motion and Special Coordinates

The equations of motion are obtained straightforwardly by varying the fields. We adopt the strategy of explicitly implementing the constraints before performing the variation. To insure that $n^\mu A_\mu = 0$, we write

$$A_\mu = n_\mu A_n + \epsilon^i_\mu A_i$$

where the $\epsilon^i_\mu$, $i = 1,2$ are a pair of mutually orthogonal spacelike vectors that are also orthogonal to $n^\mu$, and we vary $A_n$ and $A_i$ independently. Actually it's easy to see that $A_n$ drops out of the Lagrangian, so the only relevant variables are the $A_i$. We do the same for $B_\mu$.

We proceed by choosing our coordinates so that $n^\mu = (1,0,0,1)$. Our equations will no longer have manifest VSR invariance, but if we pick a different form for $n^\mu$, the VSR invariance of $L$ will insure that the end results for physically meaningful quantities will be the same.

Having chosen $n^\mu$, we define light-cone coordinates

$$u = n \cdot x = t - z, \quad \text{and} \quad v = t + z.$$

$u$ will play the role of the evolution parameter. We note that $n \cdot \partial = 2\partial_v$.

The constraint $n \cdot \partial \phi = 1$ can be satisfied by writing

$$\phi = \frac{1}{2} v + h(u, x, y).$$

Variation of $\phi$ subject to the constraint is thus tantamount to varying with respect to $h$. We shall comment on the significance of the $h$ equation in the discussion and addendum below.

To analyze the equations of motion, it is convenient to define what we call "special coordinates", which are given by

$$u' = u; \quad \vec{x}'_\perp = \vec{x}_\perp; \quad v' = v + 2h.$$

In these coordinates, the covariant derivative takes the simple form

$$D_\mu = (\partial_{u'}, \partial_{x'}, \partial_{y'}, -\partial_{u'}).$$

We have eliminated the dynamical variable $h$ by absorbing it in the definition of one of the coordinates. Indeed the coordinate $v'$ is just $2\phi$. Henceforth, we shall work in the special coordinate system, and drop the primes, so that unprimed variables will refer to the special coordinates unless otherwise specified.

We use the Bjorken-Drell convention for the $\gamma$-matrices, i.e.

$$\gamma^0 = \begin{pmatrix} I & 0 \\ 0 & -I \end{pmatrix} ; \vec{\gamma} = \begin{pmatrix} 0 & \vec{\sigma} \\ -\vec{\sigma} & 0 \end{pmatrix}.$$

As in previous papers, we expand the Fermi fields in the following basis:

$$|b_1> = \begin{pmatrix} \uparrow \\ \uparrow \end{pmatrix}; |b_2> = \begin{pmatrix} \downarrow \\ \downarrow \end{pmatrix}; |b_3> = \begin{pmatrix} \uparrow \\ -\uparrow \end{pmatrix} ; |b_4> = \begin{pmatrix} \downarrow \\ -\downarrow \end{pmatrix},$$

where $\uparrow$ and $\downarrow$ denote eigenstates of $\sigma_z$ with eigenvalues +1 and -1 respectively. $|b_1>$ and $|b_4>$ are annihilated by $n \cdot \gamma$. $|b_1>$ and $|b_2>$ are eigenstates of $\gamma_5$ with eigenvalue +1, whereas $|b_3>$ and $|b_4>$ are eigenstates of $\gamma_5$ with eigenvalue $-1$.

We write

$$\psi^{(a,b)} = \sum_{j=1}^{4} f_j^{(a,b)} |b_j>$$

with similar expansions for $\rho$ and $\chi$. We also introduce the combinations

$$C = iA_1 - A_2 + B_1 + iB_2 \text{ and } D = iA_1 + A_2 + B_1 - iB_2$$

and their complex conjugates. Since the 0 and 3 components of $A_\mu$ and $B_\mu$ do not enter the Lagrangian, we can replace $A_\mu$ and $B_\mu$ in the Lagrangian by the linear combinations $C, D, C^\dagger$ and $D^\dagger$.

**Constructing the Avatar**

The equations of motion coming from variation of $f_1^{(a)*}$ and $f_4^{(a)*}$ require $f_2^{(a)}$ and $f_3^{(a)}$ to be independent of the transverse coordinates x and y, depending therefore only on $u$ and $v$. Since these functions couple to the rest of the fermion coefficients, we make the fundamental assumption that solutions of the equations of motion are such that the entire set of coefficients depend only on $u$ and $v$. The avatar then embodies the dynamics of this two-dimensional system.

Once this assumption has been made, we find that the degrees of freedom can be assembled into two groups, which we call the $\mathcal{A}$ team and the $\mathcal{B}$ team. Members of the $\mathcal{A}$ team are $f_2^{(a,b)}, f_3^{(a,b)}, f_2^{(\chi)}, f_3^{(\chi)}, f_1^{(\rho)}, f_4^{(\rho)}, C$ and their complex conjugates. The $\mathcal{B}$ team is everything else.

The two teams are decoupled from each other. Derivatives with respect to the evolution parameter $u$ occur only in the equations for the $\mathcal{A}$ team. We therefore label the $\mathcal{A}$ team as *dynamical*, and the $\mathcal{B}$ team as *non-dynamical*. Moreover, the gauge field $C$ couples to the other members of the $\mathcal{A}$ team, but its counterpart on the $\mathcal{B}$ team, the gauge field $D$, is decoupled from all other degrees of freedom.

Given this situation, we send the $\mathcal{B}$ team to the showers, and work only with the $\mathcal{A}$ team in constructing the avatar. At the Lagrangian level,

$$L = L_\mathcal{A} + L_\mathcal{B},$$

where $L_\mathcal{A}$ and $L_\mathcal{B}$ contain only fields from their respective teams, where derivatives with respect to $u$ appear only in $L_\mathcal{A}$, and where the gauge fields couple, through the combinations $C$ and $C^\dagger$, only to the fermions in $L_\mathcal{A}$. The Lagrangians $L_\mathcal{A}$ and $L_\mathcal{B}$ describe separate systems, and it is the $\mathcal{A}$ system that contains the dynamics that will be captured by the avatar. We therefore discard $L_\mathcal{B}$ and work with $L_\mathcal{A}$.

Before proceeding, we note that there is one casualty of dropping $L_\mathcal{B}$. The symmetry that we gauged by introducing $L_{int}$ in paper III was defined by

$$\psi^{(a,b)} \to \mathcal{M}(\alpha,\beta)\psi^{(a,b)} \; ; \; \chi \to \mathcal{M}(\alpha,\beta)\chi \; ; \; \text{and} \; \rho \to \mathcal{M}(-\alpha,\beta)\rho,$$

where $\mathcal{M}(\alpha,\beta) = \exp\{\alpha n_\mu \gamma^\mu + i\beta \gamma_5 n_\mu \gamma^\mu\} = 1 + \alpha n_\mu \gamma^\mu + i\beta \gamma_5 n_\mu \gamma^\mu,$

(the last form follows because $(n \cdot \gamma)^2 = 0$). However, this symmetry, even with constant $\alpha$ and $\beta$, mixes the two teams. For example,

$$\Delta f_1^{(a,b)} = 2(\alpha + i\beta)f_3^{(a,b)}; \; \Delta f_2^{(a,b)} = 0; \; \Delta f_3^{(a,b)} = 0 \; \text{and} \; \Delta f_4^{(a,b)} = 2(\alpha - i\beta)f_2^{(a,b)},$$

with similar expressions for $\Delta f_j^{(\chi)}$ and $\Delta f_j^{(\rho)}$. So neither team exhibits the gauge invariance by itself; in constructing the avatar, we will retain the form of the gauge interaction dictated by this symmetry, but the symmetry itself will not be manifest.

Let the phases of $M_1$ and $M_2$ be $\theta$ and $\xi$:

$$M_1 = |M_1|e^{i\theta} \; \text{and} \; M_2 = |M_2|e^{i\xi}$$

and define the combinations

$$g_{2,3}^{(\pm)} = \pm e^{i(\theta-\xi)}f_{2,3}^{(a)} + f_{2,3}^{(b)}.$$

From these define the doublets

$$G^{(\pm)} = \begin{pmatrix} g_2^{(\pm)} \\ g_3^{(\pm)} \end{pmatrix}.$$

We also define

$$F^{(\chi)} = e^{-i\xi\frac{1}{\kappa}} \begin{pmatrix} f_2^{(\chi)} \\ f_3^{(\chi)} \end{pmatrix} \quad \text{and} \quad F^{(\rho)} = e^{-i\xi\frac{1}{\kappa}} \begin{pmatrix} f_4^{(\rho)} \\ f_1^{(\rho)} \end{pmatrix},$$

where $\kappa^2 = \frac{1}{2}|M_1 M_2|$, and we let

$$\mathcal{F}^{(\pm)} = |M_1| F^{(\chi)} \pm \frac{1}{2}|M_2| F^{(\rho)}.$$

In terms of these variables, which are taken to depend only on $u$ and $v$, the avatar Lagrangian $L_\mathcal{A}$ takes the form

$$L_\mathcal{A} = L_G + L_\mathcal{F} + L_M + L_{int} + L_{gauge},$$

with

$$L_G = i\left[G^{(+)\dagger}\partial_u G^{(+)} + G^{(-)\dagger}\partial_u G^{(-)}\right],$$

$$L_\mathcal{F} = i\left[\mathcal{F}^{(+)\dagger}\partial_v \mathcal{F}^{(+)} - \mathcal{F}^{(-)\dagger}\partial_v \mathcal{F}^{(-)}\right],$$

$$L_M = i\kappa\left[\mathcal{F}^{(+)\dagger}G^{(+)} - \mathcal{F}^{(-)\dagger}G^{(-)} - G^{(+)\dagger}\mathcal{F}^{(+)} + G^{(-)\dagger}\mathcal{F}^{(-)}\right],$$

$$L_{int} = g\left[G^{(+)\dagger}\begin{pmatrix} 0 & C \\ C^\dagger & 0 \end{pmatrix}G^{(+)} + G^{(-)\dagger}\begin{pmatrix} 0 & C \\ C^\dagger & 0 \end{pmatrix}G^{(-)}\right], \text{ and}$$

$$L_{gauge} = -2k\partial_v C^\dagger \partial_v C.$$

Because the variables do not carry dependence on $\vec{x}_\perp$, the part of $L_{AB}$ with the $F^2$ terms (i.e. the part proportional to the constant $K$) vanishes and hence no longer appears in $L_{gauge}$. We have defined our variables to absorb the phases of the mass parameters, showing that $\kappa$ is the only mass parameter of physical significance.

In the absence of the interaction term, the (+) piece of $L_\mathcal{A}$ separates from the (-) piece. In addition, the upper components of $G$ and $\mathcal{F}$ decouple from the lower ones. Indeed, the upper (or lower) components by themselves are equivalent to the avatar constructed in paper III. This can be seen by the following dictionary for the upper components, where the quantities from paper III are distinguished with tildes:

$$g_2^{(+)} = \tilde{g}_2^{(+)}; \; if_2^{(+)} = \tilde{g}_3^{(+)}; \; g_2^{(-)} = \tilde{g}_2^{(-)}; \; f_2^{(-)} = \tilde{g}_3^{(-)}.$$

With these substitutions, the two systems have identical Lagrangians. A similar dictionary can be constructed establishing the equivalence of the lower components with the avatar of paper III. Because of this equivalence, the quantization of the avatar in the absence of interactions proceeds exactly the same way as in paper III; the only difference is that the Fock space is the product of two identical pieces, inhabited by the upper and the lower components respectively. We refer the reader to paper III for details of this quantization, which confirms that the fields labeled with (+) describe quanta with subluminal propagation, whereas the (-) fields describe tachyons.

Unlike the avatar of paper III, $L_\mathcal{A}$, including the interaction term, is invariant under a chiral transformation:

$$G^{(\pm)} \rightarrow e^{i\alpha\sigma_z} G^{(\pm)}; \quad \mathcal{F}^{(\pm)} \rightarrow e^{i\alpha\sigma_z} \mathcal{F}^{(\pm)}; \quad C \rightarrow e^{2i\alpha} C.$$

$\sigma_z$ is the appropriate avatar of $\gamma_5$ because, for both $G^{(\pm)}$ and $\mathcal{F}^{(\pm)}$, the upper and lower components come from basis vectors with opposite eigenvalues of $\gamma_5$.

In addition, we see that, given a non-zero real constant $\lambda$, $L_\mathcal{A}$ is symmetric under $u \rightarrow \lambda u$; $v \rightarrow \lambda^{-1} v$, with the fields transforming as $G^{(\pm)} \rightarrow \lambda^{\frac{1}{2}} G^{(\pm)}$; $\mathcal{F}^{(\pm)} \rightarrow \lambda^{-\frac{1}{2}} \mathcal{F}^{(\pm)}$; $C \rightarrow \lambda^{-1} C$. This transformation, a residue of the underlying VSR symmetry, is essentially Lorentz invariance in $1+1$ dimensions.

Finally, we note that $L_\mathcal{A}$ is invariant under $v \leftrightarrow -v$; $G^{(+)} \leftrightarrow G^{(-)}$; $\mathcal{F}^{(+)} \leftrightarrow -\mathcal{F}^{(-)}$. The transformation $v \leftrightarrow -v$ is the realization, in the special coordinate system, of Light Cone Reflection, and we see that, as expected, it involves the interchange of tachyonic and non-tachyonic modes.

**Discussion**

The two-dimensional field theory defined by $L_\mathcal{A}$ bears a certain family resemblance to the massless Thirring model, which was shown long ago to be exactly solvable [7]. In particular, both models have a conserved chiral current (notwithstanding the appearance of a mass parameter in $L_\mathcal{A}$), as well as a conserved vector current that encodes fermion number conservation. Our model is more complicated, having more degrees of freedom and, if we solve for $C$, the 4-fermion coupling will be non-local, but one might still investigate whether it is exactly solvable, or at least whether partial exact results can be obtained. This is a possible direction for future work.

Our model so far exists in isolation, taking no account of the way in which neutrinos couple to the rest of the Standard Model. Here we do not offer any specific proposals, except to remark that a convenient way to couple the fermions of this model to other fields would be through the vectors $A_\mu$ and $B_\mu$, because they interact only with the dynamical $\mathcal{A}$-team fermions, so the non-dynamical sector, whose evolution is not determined by the equations of motion, will remain uncoupled.

We still have to consider the equation of motion that follows from varying $\phi$ in the original 4-dimensional action. More precisely, because $n \cdot \partial \phi = 1$, the independent variable is $h$, where $\phi = \frac{1}{2} v + h(u, x, y)$. Because $h$, and its variation, are constrained to be independent of $v$, the equation that results from varying the action will be integrated on $v$. We examined this condition in paper II for the non-interacting case. There, we argued that, for fermion fields depending only on the two variables $u$ and $\phi = \frac{1}{2} v + h(u, x, y)$, the condition is automatically satisfied. In the more general interacting case, since $L_{int}$ is independent of $\phi$, the only new feature is the presence of the term $L_{AB}$. Using the properties

$$n^\mu A_\mu = n^\mu B_\mu = 0 \text{ and } n^\mu D_\mu = 0,$$

we see that $F^A_{\mu\nu}$ in $L_{AB}$ is reduced to $F^A_{ij} = \nabla_i A_j - \nabla_j A_i$, where the indices run over $i, j = 1, 2$, and where $\nabla_i = \partial_i - (\partial_i h) n \cdot \partial$. The same applies to $F^B_{\mu\nu}$. But the assumption that $A_i$ and $B_i$ depend on $x$ and $y$ only through $\phi$ tells us that

$$\nabla_i A_j = \nabla_i B_j = 0,$$

so $L_{AB}$ vanishes, as does its variation with respect to $h$. Hence we obtain no additional information. Similar reasoning led us to omit $L_{AB}$ from the avatar Lagrangian.

Nevertheless, there ultimately has to be some way of determining $\phi$, because it governs the transformation between the special coordinate system, in which the dynamics of our model is described by the avatar, and the original spacetime coordinates $x^\mu$. Indeed, the entire dependence of the fields on the transverse coordinates $x$ and $y$ is provided by $h(u, x, y)$.

Because $D_\mu \phi = 0$ and $n \cdot \partial \phi = 1$, there are no obvious ingredients with which to construct a kinetic term for $\phi$ that respects LCR and VSR invariance. How to extend, or otherwise modify, the model to pin down $\phi$ thus remains something of a mystery.

**Addendum**: Further remarks on the $h$ problem.

We seek to construct a dynamical principle (e.g. an action or at least an equation of motion) for the field $h(u, x, y)$, which, as explained above, enters the theory when solving the constraint $n \cdot \partial \phi = 1$:

$$\phi = \frac{1}{2} v + h(u, x, y).$$

We want the equation for $h$ to respect the VSR and LCR symmetries of the theory (the chiral symmetry does not act on $h$ ).

When expressed in terms of their action on $h$, the VSR symmetries are:

--translation invariance in $u, x, y$ ;

--shift symmetry $h \to h + \kappa$ for real $\kappa$ (this comes from translation symmetry in $v$);

--rotation symmetry in the $(x, y)$ plane ;

--invariance under $h \to \lambda\, h(\lambda u, x, y)$ for real $\lambda$ ;

--invariance under

$$h(u, x, y) \to h(u, x - \alpha u, y - \beta u) - \alpha x - \beta y ,$$

for real, infinitesimal $\alpha, \beta$.

In addition, the LCR symmetry acts on $h$ as

$$h(u, x, y) \to h' = -h(u, x, y) - \frac{x^2 + y^2}{u} .$$

Note that LCR is a discrete symmetry whose square is the identity.

We consider the expression $\partial^\mu \phi \partial_\mu \phi$, which, written in terms of $h$ is

$$F(u, x, y) = 2\partial_u h - \vec{\nabla}_\perp h \cdot \vec{\nabla}_\perp h .$$

This is symmetric under all the VSR transformations except the scaling transformation $h \to \lambda\, h(\lambda u, x, y)$, under which it scales as

$$F(u, x, y) \to \lambda^2 F(\lambda u, x, y) .$$

Hence the equation $F(u, x, y) = 0$ possesses all the required VSR symmetries.

We note that $F(u, x, y)$ can also be given a geometric interpretation. If we consider a surface defined by $\phi = const.$, the induced metric, expressed in coordinates $(u, x, y)$, is

$$g_{ij} = \begin{pmatrix} -2h_u & -h_x & -h_y \\ -h_x & -1 & 0 \\ -h_y & 0 & -1 \end{pmatrix},$$

and det $g = -2h_u + h_x^2 + h_y^2 = -F(u, x, y)$. [8] (The subscripts denote differentiation with respect to the indicated coordinate).

Under LCR, $F$ transforms as

$$F \to F' = F(h') = -2\partial_u h - \vec{\nabla}_\perp h \cdot \vec{\nabla}_\perp h - \left\{\frac{4}{u}\vec{x}_\perp \cdot \vec{\nabla}_\perp h + 2\frac{x_\perp^2}{u^2}\right\}.$$

If we use the equation $F = 0$, we can recast this as

$$F' = -2\left\{\left(h_x + \frac{x}{u}\right)^2 + \left(h_y + \frac{y}{u}\right)^2\right\}$$

So in general, LCR will not take a solution of $F = 0$ into another solution, i.e. the equation $F = 0$ is not LCR symmetric. However, it does admit a particular solution,

$$h_0(u, x, y) = -\frac{1}{2}\left(\frac{x^2+y^2}{u}\right),$$

which is the unique $h$ that is LCR invariant:

$$h_0 = h_0' = -h_0 - \frac{x^2+y^2}{u}.$$

Also, we see from the form of $F'$ that, up to an additive constant, $h_0$ is the unique solution of both $F = 0$ and $F' = 0$.

These circumstances prompt us to conjecture that the proper way to proceed is to take $h = h_0$. Or, rather, in order to preserve translation invariance, we should take $h$ to be a slightly more general solution of $F = 0$,

$$h(u, x, y; a, b, c, d) = -\frac{1}{2}\left(\frac{(x-a)^2+(y-b)^2}{u-c}\right) - d.$$

In constructing the quantum theory, the constants $(a, b, c, d)$ should be treated as collective coordinates.

### Acknowledgments

We are happy to acknowledge conversations and correspondence with Robert Ehrlich, David Fairlie and Marek Radzikowski.

### References

[1] A. Chodos, **arXiv:1206.5974** , **1511.06745** , **1603.07053** , referred to as papers I, II, and III, respectively.